\documentclass{article}

\usepackage{graphics,latexsym}
\usepackage{graphicx}
\usepackage{amsmath, amssymb,amsthm}

\begin{document}

\title{The Fractal Structure of Matter \\ and the Casimir Effect}
\author{Daniele Funaro}

\date{}
\maketitle

\centerline{Department of Mathematics} \centerline{University of
Modena and Reggio Emilia}
\centerline{Via Campi 213/B, 41125
Modena (Italy)} \centerline{daniele.funaro@unimore.it}

\begin{abstract}
The zero-point radiation is an electromagnetic form of energy
pervading the universe. Its existence is granted by standard
quantum theories. We provide here an explanation based on
deterministic classical electrodynamics, by associating to
particles and nuclei a series of shells, made of constrained
photons, with frequencies decaying with the distance. Such photons
are part of a pre-existing background, evolving in vacuum even at
zero temperature, and are captured by stable subatomic particles
to form very distinctive quantized patterns. The evolving shells
bring, for instance, to the creation of a fractal-type structure
of electromagnetic layers around a conductive body. This property
is then used to justify, both qualitatively and quantitatively,
the attractive Casimir force of two metal plates. The analysis is
carried out by standard arguments, except that here the
surrounding zero-point energy is finite and, albeit with a very
complicated appearance, very well-organized.
\end{abstract}

\par\smallskip


\renewcommand{\theequation}{\thesection.\arabic{equation}}

\par\medskip
\setcounter{equation}{0}
\section{A huge electromagnetic background}

The results of this paper are consequential to those collected in
\cite{funarol}, where a throughout revision of the theory of
electromagnetism in vacuum has been presented. In that book, a set
of model equations, combining classical electromagnetism with
Euler's equation for (immaterial) fluids, are discussed. The new
model allows for the treatment of an immense variety of
electromagnetic phenomena, by prescribing the exact evolution of
the fields in a very wide context. The theory is imbedded in the
general relativity framework with the help of Einstein's equation.
In this way, the local metric turns out to be both cause and
effect in the dynamical process. As a matter of fact, the geometry
is generated by the electromagnetic environment, but, at the same
time, it drives the wave itself along geodesics paths, giving
stability to such a coupled system. We review very quickly the
main consequences of this approach and we report at the end of
this section the equations, as they are formulated in a flat
metric space. Of course, the reader interested to know more about
the model, both quantitatively and qualitatively, can find the
full set of equations and the discussion of their properties in
\cite{funarol}.

\par\smallskip

A first interesting subset of solutions that can be examined is
the one containing {\sl free-waves}. These are solitary travelling
waves, evolving according to the standard rules of geometrical
optics. Unperturbed photons with compact support and perfect
spherical waves belong for instance to this category. The
space-time geometry is mildly modified in this case and has no
influence on the natural development of the wave-fronts. The most
of the electromagnetic emissions belong however to the class of
{\sl constrained waves}, in which the interaction of different
solitons is allowed.
\par\smallskip

The simplest configuration associated with constrained waves is
the energetically isolated system consisting of two or more
rotating photons, producing a dynamic configuration in stable
equilibrium. Such a motion generates a distortion of the
space-time, via Einstein's equation, and the arrangement is
expected to be stable when the photons actually follow the
geodesics of the so deformed geometry. Explicit  solutions can be
computed in the 2-D case (see also the movies in \cite{funarow}).
At each point, the electric vector field oscillates, communicating
the vibrations to the neighboring points. The global information
travels along circles giving the idea of a rotating disk. Such
turning photons can also carry a charge and generate a
gravitational setting as a consequence of the deformation of the
geometry. The frequency of rotation is inversely proportional to
the magnitude of the disk (according to the intuition that small
bodies are related to high frequencies, and viceversa).
\par\smallskip

These objects have a finite measure, which is dictated by the
compensation properties of the space-time geometry. In fact, if
one imposes that photons must travel at the speed of light also in
a rotating setting, the clock has to be continuously adjusted as
it moves radially. One can verify from the available exact
solutions that, for a certain radius, the oscillations of the
electric field ${\bf E}$ (see  (\ref{eq:sfem1})-(\ref{eq:slor1}))
are lined up with the direction of motion ${\bf V}$ of the wave.
This is the action limit of the metric and, compatibly with the
model equations, it can be taken as a physical boundary. Going
beyond such a boundary would bring to a peripheral velocity too
large to be compensated by the metric. This is a first sign of
quantization, although quantum phenomena are not directly
imprinted in the model equations, but they only show up in
presence of special solutions (such as the ones we are now
discussing).
\par\smallskip

The 3-D version of the rotating photons displays a toroid shape
and has a straightforward relation with fluid dynamics vortex
rings, which are known to be very stable entities. For this
reason, in \cite{funarol}, the possibility that stable elementary
particles (such as the electron) could be made of rotating waves,
constrained in tiny regions of space, has been taken into
consideration. We do not insist further on this aspect, since here
it is not relevant for the developments we have in mind. We are
more concerned with a secondary effect: when some electromagnetic
energy is present around a spinning particle (whatever its
structure is), a sequence of layers may develop. Then, the core of
the particle turns out to be surrounded by numerous concentric
shells, consisting of photons rotating at different frequencies.
The fluid dynamics equivalent is a  kind of tornado (see
\cite{chinosi}, in order to have an idea on how this can be
actually realized in a 3-D environment). The angular velocity
diminishes as the respective shell becomes larger and larger.
Again, the diameter of the shells is controlled by the space-time
metric. This quantum-like mechanism, allows for the diffusion of
the electric charge to the whole space (see also section 5).
\par\smallskip

\begin{center}
\begin{figure}[!t]
\centerline{\includegraphics[width=6.2cm,height=6.2cm]{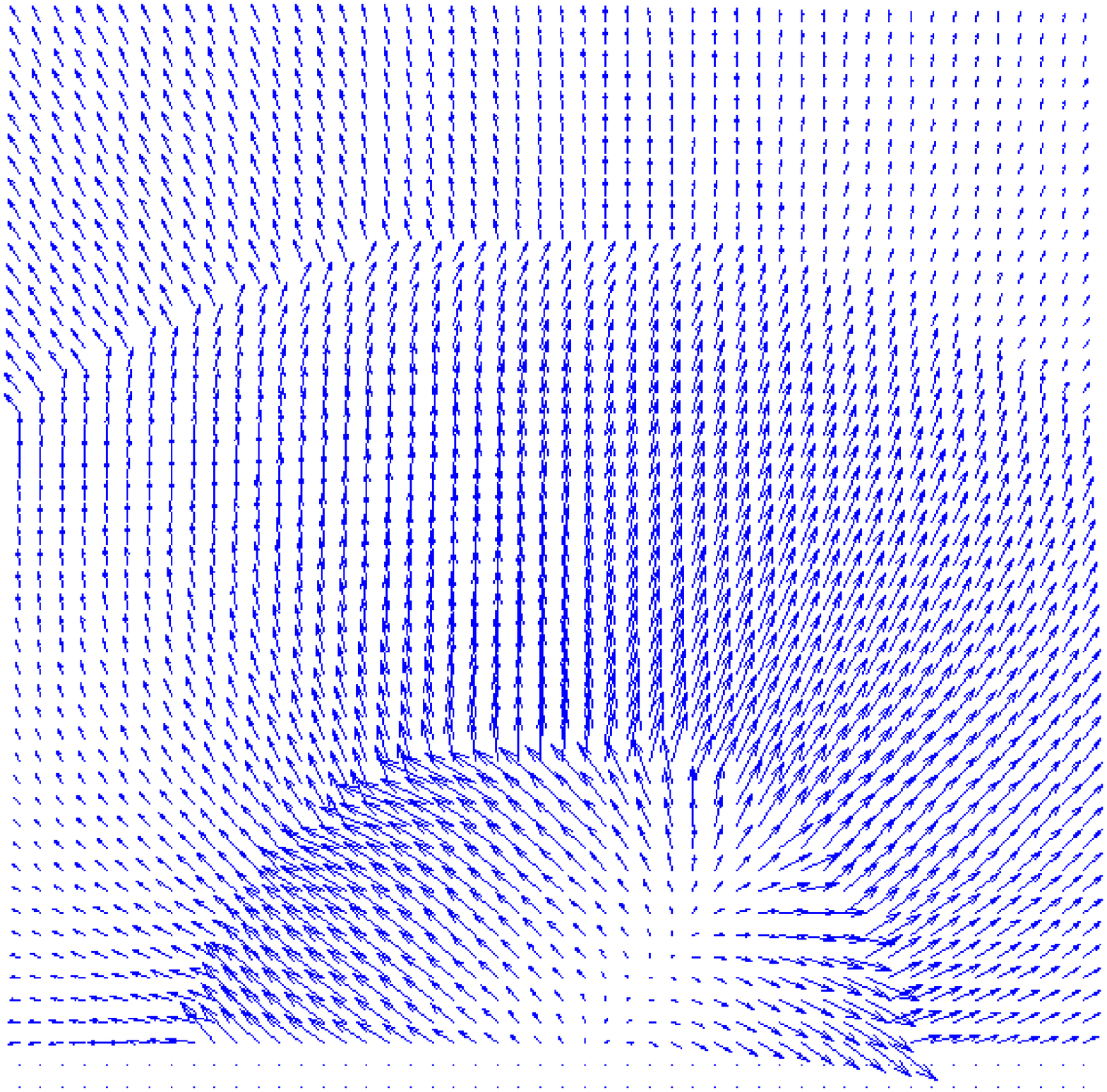}
\hspace{.2cm}\includegraphics[width=6.2cm,height=6.2cm]{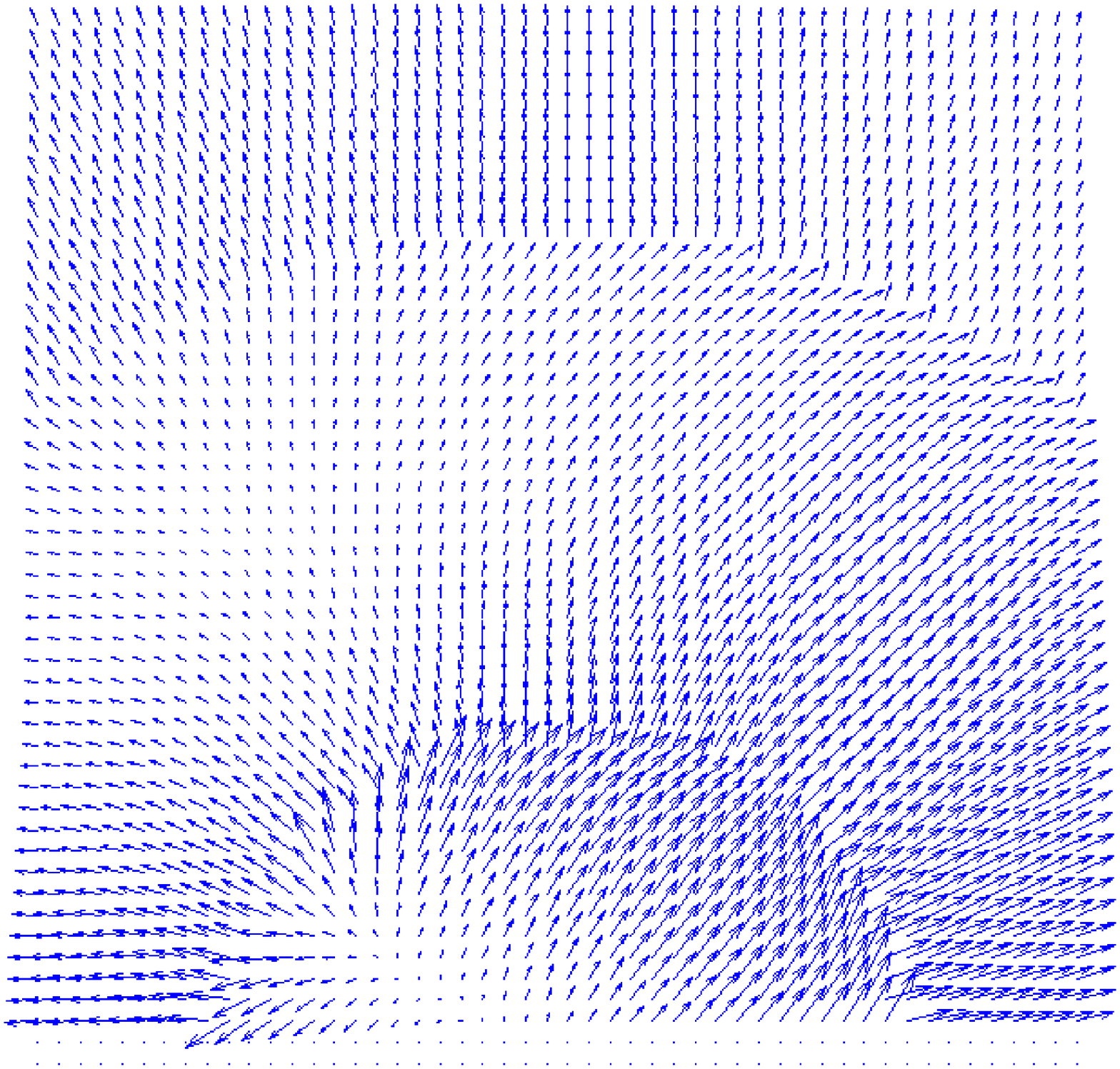}}
\begin{caption}{\small Quantized electric vector fields around a 2-D elementary
particle at different phases of evolution. The information travels
clockwise, but the angular velocity is lower as one moves from the
center. In this way, the angular momentum is preserved.}
\end{caption}
\end{figure}
\end{center}
\vspace{-1.cm}

We provide a qualitative example in figure 1 (see also the
animation in \cite{funarow}), where  part of a central kernel and
two contiguous shells are oscillating at different regimes. The
displayed vectors are related to the electric field ${\bf E}$ (the
magnetic field ${\bf B}$ is orthogonal to the page). With the
exception of the most inner part, having nonzero constant charge
density, the intensity of the vectors decays as the square of the
distance from the center, while they oscillate with a frequency
that reduces with the distance. More precisely, there is a radial
stationary continuous component decaying at infinity (as
classically assumed), perturbed by quantized zero-average
oscillations. The vectors do not actually translate, but they
carry information along closed orbits following the geodesics of a
curved space-time. Based on the results in \cite{funarol} the
movement is necessary in order to ``activate'' the tensor at the
right-hand side of the Einstein's equation, so that producing a
non flat geometry (associated to the same curved geodesics
followed by the photons). It can be shown in fact that stationary
solutions do not deform the geometry in the appropriate way and
bring to unwanted point-wise singularities.
\par\smallskip

Two oscillating layers are divided by a separation surface and the
information travels at different speeds on both sides.
Nevertheless, no discontinuities are actually measured. First of
all, according to the model equations (see (\ref{eq:slor1})), this
is true because the velocities, independently on both sides,
follow the rules of inviscid fluid dynamics. Secondly, the
space-time geometry is deformed in such a way that, after writing
the electromagnetic constitutive equations in the new metric,
there are no derivatives to be computed in a direction transversal
to the separation surface. The central core and the successive
shells are independent entities that do not communicate directly,
but only through slight vibrations of the separation surfaces,
around the position of equilibrium (as it happens in any perturbed
mechanical system). Discontinuities are consequence of the wrong
idea that such a complicated phenomenon can be actually explained
from the framework of a flat geometry. A characterization of a
separation surface is obtained by observing that the derivative
with respect to time of the electric field is tangent to it, i.e.,
the time-dependent component of the field, orthogonal
to the surface, is zero. This brings to a local collapse of the metric and
the generation of a successive shell, if further energy is
available. Such an observation will be useful later, when we
prescribe a way to detect the separation surfaces without
computing the entire field.
\par\smallskip

Before ending this section, we quickly introduce the equations studied in \cite{funarol}. In a
flat metric space, they read as follows:
\begin{equation}\label{eq:sfem1}
\frac{\partial {\bf E}}{\partial t}~=~ c^2{\rm curl} {\bf B}~
-~\rho {\bf V}
\end{equation}
\begin{equation}\label{eq:sfbm1}
\frac{\partial {\bf B}}{\partial t}~=~ -{\rm curl} {\bf E}
\end{equation}
\begin{equation}\label{eq:sfdb1}
{\rm div}{\bf B} ~=~0
\end{equation}
\begin{equation}\label{eq:slor1}
\rho\left(\frac{D{\bf V}}{Dt}~+~\mu ({\bf E}+{\bf V}\times {\bf
B}) \right) ~=~-\nabla p
\end{equation}
with $\rho ={\rm div}{\bf E}$, where ${\bf E}=(E_1,E_2,E_3)$ is
the standard electric field and ${\bf B}=(B_1,B_2,B_3)$ the
magnetic field. Moreover ${\bf V}=(V_1,V_2,V_3)$ is a velocity
field and the triplet $({\bf E},{\bf B},{\bf V})$ is right-handed.
The given constant $\mu$ is a charge divided by a mass. Both
$\rho$ (a kind of density) and $p$ (a kind of pressure) may also
take negative values. There is an additional ``equation of state''
relating  $p$ with the scalar curvature $R$ of the space-time,
however, for simplicity, we do not discuss this technical aspect
here. There is also a perfect compatibility between the model
equations and the request that the divergence of a suitable energy
tensor (obtained as the sum of the standard electromagnetic stress
tensor and a mass-type tensor with density $\rho$) must be zero.
\par\smallskip

Free-waves are modelled by setting $R=0$, $p=0$, $\frac{D}{Dt}{\bf
V}=0$ (rectilinear geodesics). In this way, equation
(\ref{eq:slor1}) becomes: ${\bf E}+{\bf V}\times {\bf B}=0$.
Equation (\ref{eq:sfem1}) turns out to be the Amp\`ere law for a
freely flowing unmaterial current with density $\rho$ (actually,
in this circumstance, the Lorentz acceleration $\frac{D}{Dt}{\bf
V}$ vanishes).
\par\smallskip

We recapitulate by saying that the evolution of electromagnetic
phenomena (in principle of any kind) is governed by deterministic
time-dependent equations (a nonlinear version of Maxwell's
equations). The light rays can be interpreted as stream-lines of a
fluid described by the non viscous Euler equations. These lines
are associated with the geodesics of a space-time geometry. In
synchronism, the geometry is perturbed by the passage of the wave.
Therefore, every electromagnetic phenomenon is unavoidably related
to a dynamical gravitational setting. The set of equations
fulfills all the basic assumptions of invariance classically
required in physics. Moreover, we could potentially build a model
for describing  matter by using exclusively electromagnetic waves.
\par\smallskip

This synthetic introduction may sound a little extravagant.
Nevertheless, let us start with these pre-requisites to see if it
is possible to translate these ideas in a mathematical language
and provide alternative explanations to known facts, such as the
Casimir effect.

\par\medskip
\setcounter{equation}{0}
\section{Some quantitative results}

According to the scenario depicted in the previous section, matter
is made of massive subatomic particles only in small amount. The
rest is electromagnetic energy organized by these particles to
follow specific patterns. Due to its oscillating nature, such a
background radiation is almost invisible, but it can be easily
revealed through indirect experiments (for example, every time we
observe photon emission, it means that a shell of a certain
frequency has been released). The solution of the entire system of
equations proposed in \cite{funarol} should be able to provide all
the necessary details to understand the structure of complicated
molecular aggregations (for what concerns the evolution of the
electromagnetic background, at least). Of course, this is quite an
impressive amount of work, therefore, the search of a simplified
model is a compulsory step.
\par\smallskip

For the applications  we have in mind, it will be not necessary to
know the exact evolution of the electromagnetic fields. It is
sufficient instead to have an idea of the location and the size of
the various shells and the frequency they carry. To this end, we
will combine two hypothesis: the oscillating nature of the scalar
electric potential and the inverse linear decay of the frequency
rate as a function of the distance from the source.
\par\smallskip

We denote by $\Phi$ the scalar electric potential and, as usual,
we assume it satisfies the following wave equation:
\begin{equation}\label{eq:onde}
\frac{1}{c^2}\frac{\partial^2\Phi}{\partial t^2}~=~\Delta \Phi
\end{equation}
where $c$ denotes the speed of light.
\par\smallskip

We do not need to know exactly what is an electron (or a positive
nucleus). We just assume that it is something with a diameter
$\delta >0$ and imbedded in a sequence of spherical shells. Each
shell is made of rotating photons, associated with a frequency:
\begin{equation}\label{eq:freqde}
\nu~\approx~\frac{c\beta}{r}
\end{equation}
where $r>\delta$ indicates the distance from the source and
$\beta$ is a given constant. In truth, the frequency should
manifest a series of jumps, therefore, it is not a continuous
function of $r$ as indicated in (\ref{eq:freqde}). The problem is
that, for the moment, we do not know the location of these jumps.
By using the uniformly spherical expression (\ref{eq:freqde}), we
also lose the notion of spin; however this is not a crucial aspect
here.
\par\smallskip

According to the above hypotheses, we would like to get rid of the
time variable in (\ref{eq:onde}). If $\Phi$ oscillates as $~\sin
(\nu t)$, we substitute the second derivative in time by:
$~-\nu^2\sin (\nu t)$. Successively, we replace $\nu$ by the
expression in (\ref{eq:freqde}). These passages suggest to
introduce a new function $\tilde\Phi$, that will be required to
satisfy the stationary equation:
\begin{equation}\label{eq:lapote}
\Delta \tilde\Phi~+~\frac{\beta^2}{r^2}\tilde\Phi~=~0
\end{equation}
It should be clear that there is no direct relation between $\Phi$
and $\tilde\Phi$, since by applying the $\Delta$ operator we did
not compute the spatial derivatives of $\nu$. The equation
(\ref{eq:lapote}) is  however of primary importance for the
discussion to follow.
\par\smallskip

We know that the gradient of the function $\tilde\Phi$ is
orthogonal to its level surfaces. If, in addition, these level
surfaces consist entirely of stationary points, the vector
$\nabla\tilde\Phi$ is identically zero there. Roughly speaking,
forgetting the time dependency, we have that that
$\nabla\Phi\approx\nabla\tilde\Phi$  is the electric field, and we
would like to find, at least qualitatively, the surfaces where
$\nabla\Phi$ has no component orthogonal to the local tangent
plane. As we said in section 1, these are transition surfaces
between one shell and the next one.
\par\smallskip

The conclusions are heuristical but quite effective. For a given
solution of (\ref{eq:lapote}), we compute the corresponding
surfaces made of stationary points (as we shall see later, it will
be simpler to compute the zeros). Then, we claim that, within a
reasonable degree of approximation, these surfaces represent the
boundaries of the different oscillating shells. Note instead that
the behavior of $\tilde\Phi$ between these surfaces has no
physical meaning. Therefore, the technique we are going to explain
is only finalized to find the location of the separation surfaces.
This preliminary procedure will be then followed by other steps,
with the scope to evaluating the global energy surrounding a piece
of matter.
\par\smallskip

From the mathematical point of view, equation (\ref{eq:lapote}) is
very peculiar. The operator $\Delta$ is negative definite, so that
there is a balance with the positive term $(\beta /r)^2$. The
result is an indefinite, variable coefficients, Helmholtz problem.
We observe that (\ref{eq:lapote}) admits self-similar solutions.
As a matter of fact, a dilation of the $r$-axis does not alter the
nature of the equation, since $\Delta$ behaves as $1/r^2$. This
remark is important, because it will reveal a fractal underlying
structure.
\par\smallskip

We can write the Laplacian in (\ref{eq:lapote}) in spherical
coordinates. By looking for pure radial functions, after
separation of variables, one obtains:
\begin{equation}\label{eq:laposep}
u^{\prime\prime}~+~\frac{2}{r}u^\prime~+~\frac{\beta^2}{r^2}u~=~0
\end{equation}
admitting, up to multiplicative constants, solutions of the following form:
\begin{equation}\label{eq:solsf}
u(r)~=~\frac{1}{\sqrt{r}}~\sin\hspace{-.08cm}\left(\sqrt{\beta^2
-{\textstyle\frac{1}{4}}} ~\log r\right)~~~~~{\rm for}~~\beta >
{\textstyle\frac{1}{2}}
\end{equation}
The above solution is an eigenfunction, corresponding to the zero
eigenvalue, of the differential operator in (\ref{eq:laposep}).
Another branch of solutions is obtained by replacing the sinus by
the cosinus. Note that, by the substitution $~r\rightarrow
re^\gamma ~$ with $~\gamma =\pi /\sqrt{\beta^2-\frac{1}{4}}$, we
reobtain $u$ up to a multiplicative constant.
\par\smallskip

Concerning boundary conditions, it is reasonable to assign the
value of the electric field $\nabla\Phi\approx\nabla\tilde\Phi$.
In terms of the function $u$, we can impose the value of its
derivative at the point $r=\delta$ and require that $u\rightarrow
0$ for $r\rightarrow +\infty$. In alternative, one can follow the
solution in (\ref{eq:solsf}) up to $r=0$, discovering infinite
oscillations.
\par\smallskip

Let us finally note that, since we are only interested to describe
a qualitative behavior, it makes no difference to compute either
the zeros or the stationary points of $u$ (or the equivalent
version with the cosinus). Due to the self-similarity properties
of equation (\ref{eq:laposep}), these sets of points are
interlaced and it will be sufficient to adjust the size of
$\delta$ to pass from one set to the other.
\par\smallskip

Therefore, let us compute for instance the zeros of
(\ref{eq:solsf}). These are:
\begin{equation}\label{eq:zerir}
r_k~=~e^{\gamma k}~~~~~ k~{\rm integer},~~~~~~~~~~{\rm
with}~~\gamma=\frac{\pi}{\sqrt{\beta^2 -1/4}}
\end{equation}
implying that the amplitudes of the shells grow geometrically.
\par\smallskip

Once we have found the location of the transition boundaries, we
should assign a frequency to each shell. Due to (\ref{eq:freqde}),
this is done as follows:
\begin{equation}\label{eq:freqir}
\nu_k~=~\frac{c\beta}{r_{k+1}}~~~~~~\forall ~r\in ]r_k,r_{k+1}]
\end{equation}
where the index $k$ can be also negative.
\par\smallskip

In addition, we would like to provide each shell with a suitable
energy. Being a shell composed by photons of frequency $\nu_k$, it
is customary to introduce a ground-state energy ${\cal E}_k$ per
unit of volume in order to satisfy the following relation:
\begin{equation}\label{eq:enerir}
\int{\cal E}_k ~dV~=~4\pi \int_{r_k}^{r_{k+1}}\hspace{-.2cm}{\cal
E}_k ~r^2dr~=~{\textstyle{\frac{1}{2}}}h\nu_k
\end{equation}
where $h$ is the Planck constant. If ${\cal E}_k$ is constant on
the entire shell, we get: ${\cal E}_k\approx 1/r^4_k = e^{-4\gamma
k}$, which is quite a fast decay for $k\rightarrow +\infty$. Note
that the computation of the integral of
$\vert\nabla\tilde\Phi\vert^2$ does not provide significant
insight, since $\tilde\Phi$ (determined up to multiplicative
constant) is not the electric potential associated with the energy
distribution in (\ref{eq:enerir}). We will also assume that there
is a maximum frequency $\nu_{\rm max}=c\beta /\delta$, so that
there exists a minimum index $\hat k$ such that:
\begin{equation}\label{eq:limite}
\nu_k~<~\nu_{\rm max}~~~~~~~\forall ~k>\hat k
\end{equation}

\par\smallskip

If to the time-dependent electric field we sum up a stationary
component $~q/(4\pi\epsilon_0r^2)$, for $r>\delta$, one can
recover an interesting relation for the energy:
\begin{equation}\label{eq:staz}
4\pi\epsilon_0
\int_{r_k}^{r_{k+1}}\hspace{-.2cm}\left(\frac{q}{4\pi\epsilon_0r^2}\right)^
{\hspace{-.1cm}2}r^2dr ~=~\alpha~ \frac{e^\gamma
-1}{2\pi\beta}~h\nu_k
\end{equation}
where $q$ is the electron charge and $~\alpha =q^2/2hc\epsilon_0~$
is the fine structure constant. Thus, we got a well-known result:
the energy in (\ref{eq:staz}) is proportional to the Planck
constant multiplied by the shell frequency. Note that this is not
true with a general distribution of points $r_k$, but only when
their growth is geometrical as in (\ref{eq:zerir}). Note also that
the stationary potential $~\Phi_0 =-q/(4\pi\epsilon_0r)~$ is
solution to the wave equation (\ref{eq:onde}).

\par\medskip
\setcounter{equation}{0}
\section{Organized energy patterns}

It should be clear to the reader how important is the existence
and the characterization of some extensive electromagnetic
background, in order to explain the Casimir effect (see
\cite{casimir}, \cite{casimir2}). We remind that this phenomenon
concerns the attraction of two parallel uncharged metallic plates
in vacuum. Thus, it would be worthwhile to recall first the
official theoretical justification. This is based on the fact that
some form of energy circulating at the exterior of the plates is
larger than that trapped in between, resulting in a gradient of
pressure pushing on the surfaces. The existence of the so called
{\sl zero-point} energy is something predicted by standard quantum
mechanics (see, e.g., \cite{dodd},  \cite{milonni}, \cite{ibison},
\cite{loudon}, for a general overview). Such energy remains even
when all the usual sources are removed from the system. It is a
kind of fuzziness, attributed to matter, at a minimum uncertainty
energy level, as a consequence of the Heisenberg principle. Its
presence is however obscure in the classical non-quantum context.
\par\smallskip

The attractive force is relatively strong, since it behaves as
$d^{-4}$, where $d$ is the distance of the plates ($d$ small
enough).  More precisely, denoting by $A$ the area of the plates,
for $d<\hspace{-.1cm}<\sqrt{A}$, the force is given by:
\begin{equation}\label{eq:casa}
F~=~-\frac{hc\pi}{480~ d^4}
\end{equation}
\par\smallskip

In the theoretical analysis, the electromagnetic radiation around
the plates is represented as an infinite sum of modes with
averaged minimum energy. Then, one observes that, since the plates
are perfectly conducting and uncharged, in the space between them,
only a subset of the modes has to be summed up. Thus, the
difference of the outside and the inside energies is not zero,
producing the Casimir effect. A nontrivial trouble is due to the
fact that both the terms of the difference are divergent sums.
This problem is generally overcome with mathematical arguments,
not always clearly justified.
\par\smallskip

Our aim is twofold. First of all, we argue that the zero-point
energy can be described deterministically,  in terms of classical
electrodynamics, via the equations introduced in \cite{funarol}.
Secondly, such an energy is finite, eliminating the inconvenience
of handling divergent sums.
\par\smallskip

In order to proceed, we need to make some assumptions about the
structure of a metallic surface $\Omega$. As we said in the
previous section, electrons and nuclei carry, together with a
stationary component providing for an effective charge, a system
of electromagnetic shells vibrating at different frequencies. In
the average, at a certain distance from $\Omega$, the charges
compensate, so that we can get rid of the stationary part. We keep
instead the oscillating component. If we are extremely close to
the atoms, such a component is expected to display a very
complicated evolution. But, as we move far from the molecular
structure, one could simplify the setting and thinking of a
sequence of plane parallel layers (neglecting unpredictable
effects near the boundary of $\Omega$), each one carrying a
frequency that decays with the distance from $\Omega$. Note that
the amplitude of the shells around the atoms grows geometrically
(see (\ref{eq:zerir})). Contrary to what would happen with a
linear growth, the interference patterns, created by shell systems
associated with different atoms, rapidly fade, giving rise to
smooth flat parallel electromagnetic layers. In practice, we are
assuming some symmetry properties of the molecular lattice, which
might not be true in the case of nonconductive materials.
\par\smallskip

A way to model this behavior is to take equation (\ref{eq:lapote}),
and substitute the distance $r$ from a single particle
with the distance $x$ from $\Omega$:
\begin{equation}\label{eq:lapote1}
\Delta \tilde\Phi~+~\frac{\beta^2}{x^2}\tilde\Phi~=~0
\end{equation}
where $\tilde\Phi$ depends on the variable $x$ and the two
variables $y$ and $z$, extending on the plate surface. Although
this is not a real restriction, let us suppose for simplicity that
$\Omega$ is a square. Now, if we use separation of variables, we
arrive to:
\begin{equation}\label{eq:lapotem}
u^{\prime\prime}~+~\left(\frac{\beta^2}{x^2}-\lambda^2\right) u~=~0
\end{equation}
Here $\lambda$ is an eigenvalue of the 2-D Laplacian on the domain
$\Omega$. Indeed, there are infinite eigenvalues, that can be
related  to the area $A$ of $\Omega$ in the following way:
\begin{equation}\label{eq:eige}
\lambda~=~2\pi\sqrt{\frac{n^2+m^2}{A}} ~~~~~~n, m~{\rm
positive}~{\rm integers}
\end{equation}
The corresponding eigenfunctions, depending on the variables $y$
and $z$, have zero average in $\Omega$.
\par\smallskip

\begin{center}
\begin{figure}[!t]
\centerline{\includegraphics[width=6.2cm,height=6.2cm]{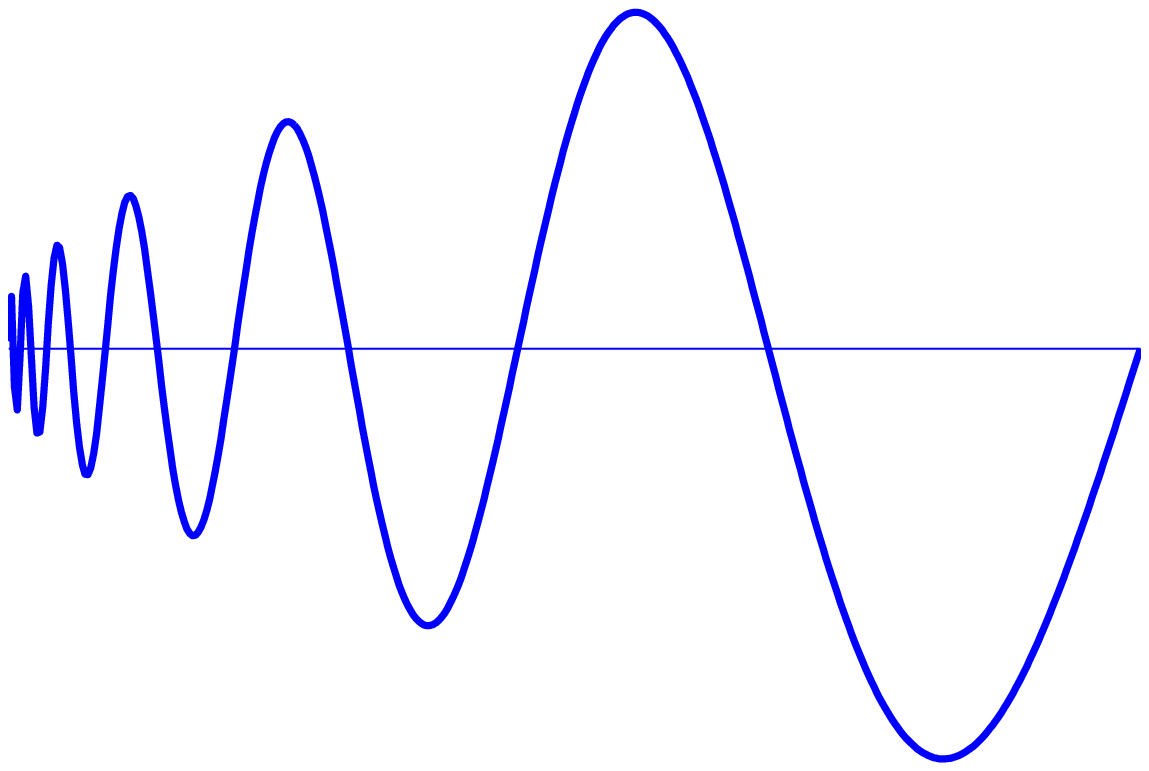}
\hspace{.2cm}\includegraphics[width=6.2cm,height=6.2cm]{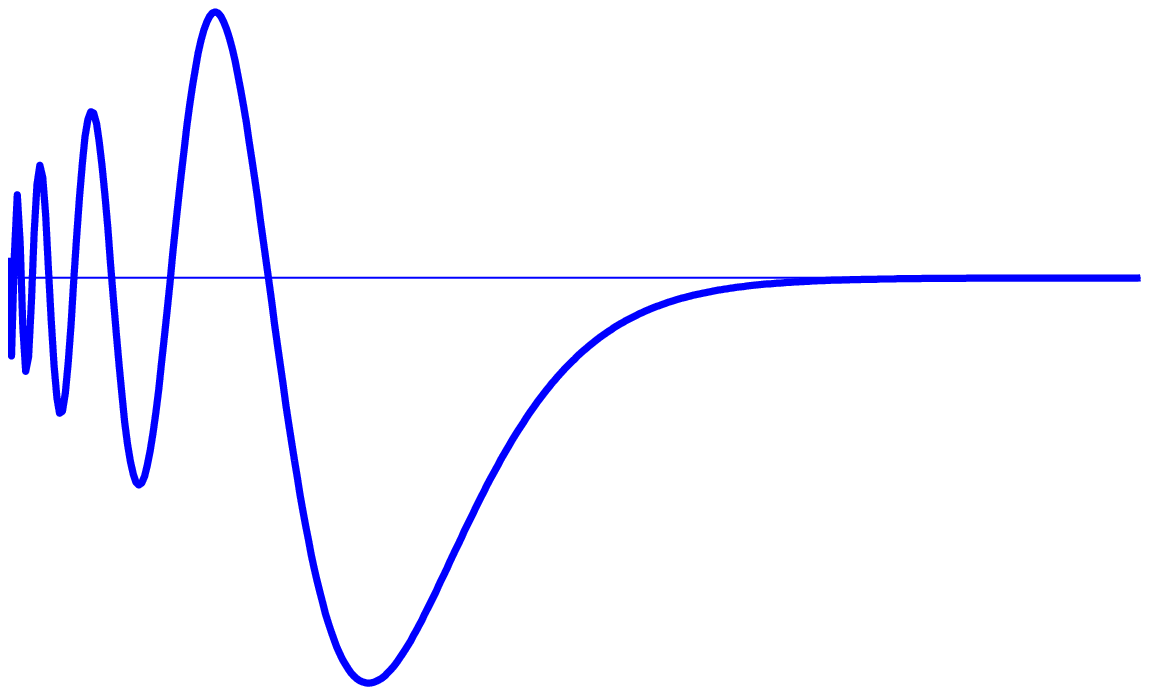}}
\begin{caption}{\small Solutions of the equation (\ref{eq:lapotem})
for $\lambda =0$ and $\lambda >0$.}
\end{caption}
\end{figure}
\end{center}
\vspace{-.5cm}

According to the recipe illustrated in section 2, the successive
step is to examine the distribution of the zeros $x_k$ of $u$. To
each interval we can then associate a frequency:
\begin{equation}\label{eq:freqi}
\nu_k~=~\frac{c\beta}{x_{k+1}}~~~~~~\forall x\in ]x_k,x_{k+1}]
\end{equation}
and a ground-state energy ${\cal E}_k$ per unit of volume:
\begin{equation}\label{eq:eneri}
A\int_{x_k}^{x_{k+1}}\hspace{-.2cm}{\cal E}_k ~dx~=~{\textstyle{\frac{1}{2}}}h\nu_k
\end{equation}

These computations are quite simple in the special case  $\lambda
=0$ (not included in (\ref{eq:eige})), where we can find an
explicit solution of (\ref{eq:lapotem}):
\begin{equation}\label{eq:sold}
u~=~\sqrt{x}~\sin\hspace{-.08cm}\left(\sqrt{\beta^2 -{\textstyle\frac{1}{4}}}
~\log x\right)
\end{equation}
The plot of (\ref{eq:sold}) is visible in the first picture of
figure 2. Another solution is recovered by replacing the sinus by
the cosinus. As we said, we are concerned with studying the
behavior of zeros, minima and maxima, of the function $u$. For
instance, by computing the zeros, we get:
\begin{equation}\label{eq:zerix}
x_k~=~e^{\gamma k}~~~~~ k~{\rm integer},~~~~~~~~~~{\rm with}~~\gamma=\frac{\pi}{\sqrt{\beta^2
-1/4}}
\end{equation}

\noindent If ${\cal E}_k$ is constant on the $k$-th shell, from
(\ref{eq:eneri}) we obtain:
\begin{equation}\label{eq:enerix}
{\cal E}_k ~=~\frac{ch\beta}{2A(1-e^{-\gamma})~{x^2_{k+1}}}
\end{equation}

\begin{center}
\begin{figure}[!h]
\centerline{\includegraphics[width=13.cm,height=4.cm]{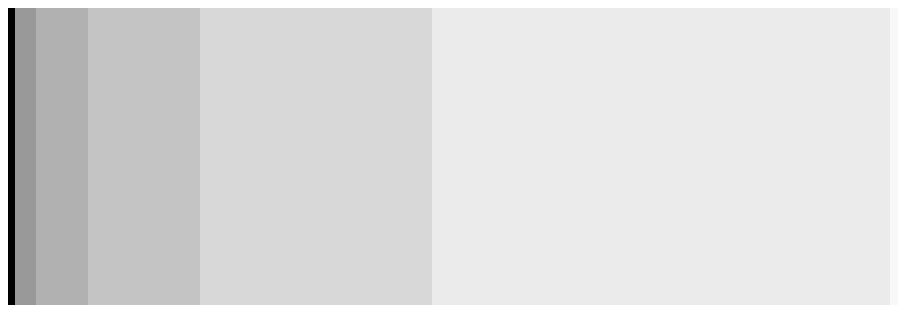}}\vspace{-.3cm}
\begin{caption}{\small Stratification of electromagnetic layers outside
a conductive surface ($\lambda =0$ in (\ref{eq:lapotem})). Each
band is related to a complex evolutive framework of
electromagnetic fields vibrating with a frequency that decays
inversely with the distance from the plate.}
\end{caption}
\end{figure}
\end{center}
\vspace{-.5cm}

The distribution of the various energy layers at the ground-state,
is qualitatively illustrated in figure 3, by different scales of
grey. With the distance the bands increase their width, but the
energy levels reduce quite fast. Now, the sum of the total energy
is finite, whenever $\delta >0$. For $\delta$ tending to zero the
sum diverges, but this occurrence has no physical justification,
since it means that we are not giving a sort of ``granularity'' to
the molecular structure. In order to stay away from the plate, we
define $\nu_{\rm max}=c\beta /\delta$ and use only frequencies
$\nu_k$ satisfying (\ref{eq:limite}).
\par\smallskip

The situation is a bit different when $\lambda >0$. In this case,
we do not have the explicit solution. However, some theoretical
considerations can be made. One can check that, in the interval
$]\delta , +\infty[$, there is only a finite number of zeros (see
the second picture of figure 2). The last inflection point of the
function $u$ is for $x=\beta /\lambda$, followed by an exponential
decay without oscillations. According to (\ref{eq:eige}), when
$n=m=1$, there are no more zeros for $x$ greater than
$\beta\sqrt{A}$. A plot of the zeros and the energy levels is
given in figure 4. Beyond the last zero, there is no appreciable
energy. We recall that we are working at the minimum temperature,
otherwise the situation would be rather different.
\par\smallskip

As a boundary condition we can require for instance that $u(\delta
)=1$ (or $u^\prime (\delta )=1$, providing a qualitatively similar
behavior). By the way, we can also impose $u(\delta )=1$ and
$u^\prime (\delta )=0$ at the same time. In this case $u$ does not
tend to zero for $x\rightarrow +\infty$, but grows exponentially.
This behavior does not affect however the qualitative displacement
of the zeros.
\par\smallskip

A rough evaluation of the zeros of $u$ is obtained by rewriting
(\ref{eq:lapotem}) in the following way:
\begin{equation}\label{eq:lapotemm}
u^{\prime\prime}~+~f^2 u~=~0~~~~~~~~~~{\rm
with}~~f=\sqrt{\frac{\beta^2}{x^2}-\lambda^2}
\end{equation}
for $x<\beta /\lambda$. We then integrate $f$ with respect to $x$
and find the values such that a primitive $F$ is equal to $k\pi$,
with $k$ integer. These passages are not rigorously justified,
but the idea is that $u$ approximately behaves as
the function $\sin (F)$.
\par\smallskip

\begin{center}
\begin{figure}[!h]\vspace{-.5cm}
\centerline{\includegraphics[width=11.6cm,height=9.cm]{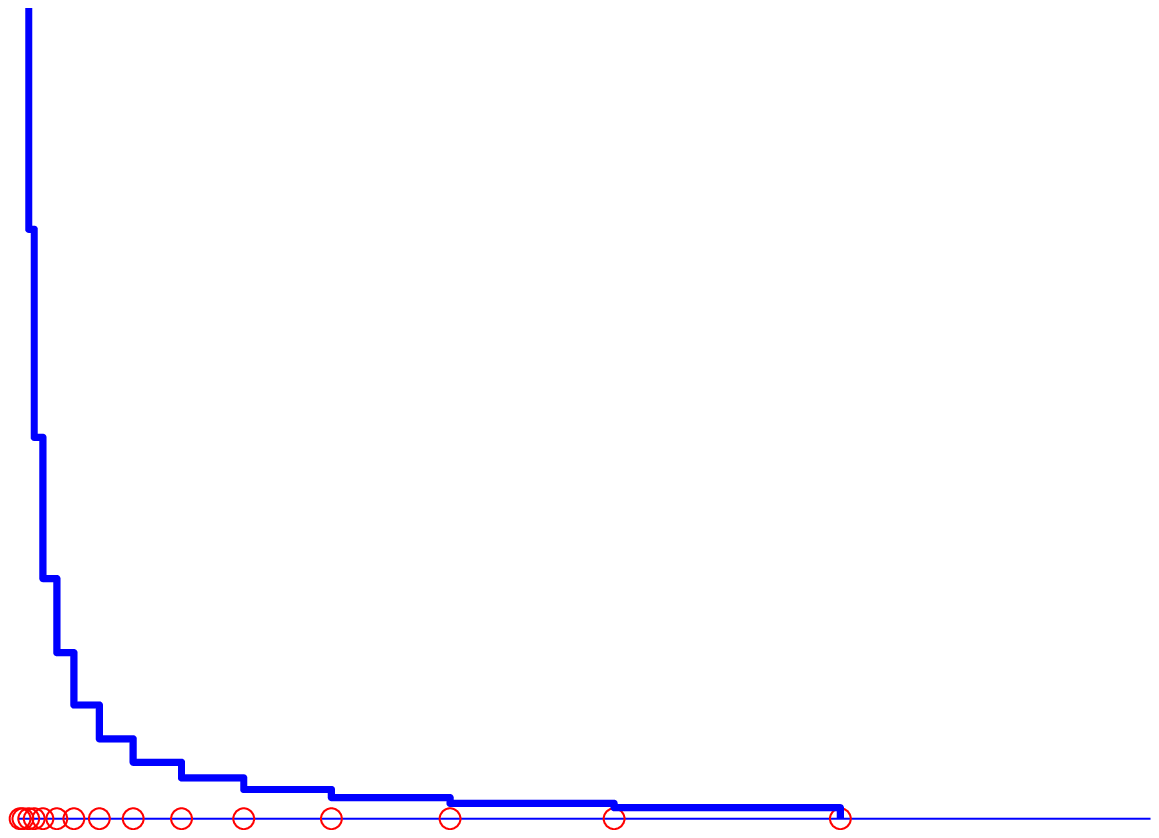}}\vspace{-1.2cm}
\begin{caption}{\small Energy levels of the layers associated to a conductive plate,
for a given $\lambda$ in (\ref{eq:lapotem}). The dots are placed correspondingly
to the zeros of $u$.}
\end{caption}
\end{figure}
\end{center}
\vspace{-.3cm}

For example, with  $f$ defined as in (\ref{eq:lapotemm}), we have:
\begin{equation}\label{eq:primi}
F~=~ x f~-~{\textstyle{\frac{1}{2}}}\beta ~\log \frac{\beta+xf
}{\beta-xf }~~~~~~~~~{\rm with}~~x<\frac{\beta}{\lambda}
\end{equation}
The integration constant is such that $F=0$ when $f=0$, and the
function $F$ turns out to be negative. When we compute the values
such that $F=k\pi$ ($k$ negative integer), we get a sequence
approaching zero with a behavior similar to that displayed by the
zeros of $u$ (see the dots of figure 4, which have been computed
numerically by solving the differential equation). The number of these points
is finite, when we are in the interval $ ]\delta, \beta /\lambda[$, for a positive
$\delta$.
\par\smallskip

From the above arguments we can draw some conclusions. We are
concerned with finding solutions $\tilde\Phi$ of
(\ref{eq:lapote1}). We require that $\tilde\Phi$ has zero average
on the square plate $\Omega$ and decays to zero far from $\Omega$.
To this end, by separation of variables, we isolate the modes
$\sin ny$, $\sin mz$, transversal to $\Omega$, from the modes
obtainable from the solution $u$ of the single variable equation
(\ref{eq:lapotem}), with $\lambda$ given in (\ref{eq:eige}).
Successively, for any fixed $\lambda$, we find the total
ground-state energy ${\cal E}_\lambda$. This corresponds to a sum
of suitable energy densities, carried by a sequence of independent
layers parallel to $\Omega$ (see figure 3). For $\delta >0$ (the
``molecular rugosity'' of the plate), ${\cal E}_\lambda$ turns out
to be finite.
\par\smallskip

The successive step is to compose the sum of the various ${\cal
E}_\lambda$, with respect to the positive integers $n$ and $m$. It
is very interesting to observe that the electromagnetic parallel
bands obtained for a certain $\lambda_1$ are exactly distributed
as those obtained for another $\lambda_2$, after performing the
linear transformation: $x\rightarrow \lambda_1 x/\lambda_2$. Once
again, this reveals a fractal structure of the electromagnetic
cloud circulating around the plate. As we said, the bands are
comprised  in the interval $ ]\delta, \beta /\lambda[$. Since
$\delta >0$, the integers $n$ and $m$ cannot be greater than a
certain amount. This means that there is a finite number of terms
${\cal E}_\lambda$. Therefore, the global energy is finite. For
$\delta\rightarrow 0$, one approaches the molecular structure of
the metallic plate, which is made of an extraordinary large (but
finite) number of atoms. The theory is coherent because, at
molecular level, there is no need to take into account the modes
with $n$ and $m$ larger than a given limit, so that the choice of
a very small (but positive) $\delta$ is well-suited to the
circumstance. In our qualitative analysis, we have no practical
suggestions about the magnitude of $\delta$. However, from the
discussion above, it should be clear that its determination must
be related to the interatomic distance (about one \AA ngstrom).
\par\smallskip

Thus, according to our model, the electromagnetic background, at
its minimum energy level, is organized by a metallic body to form
overlapped sequences of parallel films. Each sequence displays a
geometrical growth and all the sequences can be transformed one
into another by means of linear contractions. This construction
may in part explain the fractal structure of matter (see for
instance \cite{mandelbrot} or \cite{peitgen}). Our aggregations
develop in the orthogonal direction to the surface and, more
realistically, they are also time-dependent. It is not excluded
that they may contribute to the formation of fractal stencils
spread horizontally on the surface, as it is observed in a lot of
applications, such as the analysis of fracture behavior.
Fractal formations in the micro-world have been documented in a lot
of circumstances, such as in the growth of electrochemical deposition.
Among the numerous articles, we just mention a few papers:
\cite{barabasi}, \cite{bradley}, \cite{marani}.

\par\medskip
\setcounter{equation}{0}
\section{An explanation of the Casimir effect}

The aim of the previous section was to show that the
electromagnetic energy (always present, even at zero temperature),
outside a piece of conductor, is distributed according to
well-established patterns. Moreover, its total amount is finite.
To explain the Casimir effect, we can now proceed with the usual
arguments, that is, there is less energy trapped between two
plates than outside.
\par\smallskip

In order to handle the case of two parallel plates at a distance
$d$, we modify (\ref{eq:lapotem}) as follows:
\begin{equation}\label{eq:lapotem2}
u^{\prime\prime}~+~f^2 u~=~0~~~~~~~~{\rm
with}~~~~~f=\sqrt{\beta^2\left(\frac{1}{x^2}+\frac{1}{(d-x)^2}\right)-\lambda^2}
\end{equation}
Now $f$ is singular both for $x=0$ and $x=d$.  The expression of
$\lambda$ is given in (\ref{eq:eige}). The argument of the square
root is positive when $d$ is sufficiently small. To this purpose,
for $n=m=1$, the condition $d<\beta\sqrt{A}~$ is good enough. The
other cases will be discussed later.
\par\smallskip

Other choices may be admissible for $f$, such as, for instance:
\begin{equation}\label{eq:funzf}
~f=\sqrt{\beta^2\left(\frac{1}{x}+\frac{1}{d-x}\right)^{\hspace{-.1cm}2}-\lambda^2}
~~~~{\rm or}~~~~~
f=\sqrt{\left(\beta\max\left\{\frac{1}{x},\frac{1}{d-x}\right\}\right)^{\hspace{-.1cm}2}-\lambda^2}
\end{equation}
depending on how we measure the distance from the plates.
They bring, more or less, to the same results.
\par\smallskip

Afterwards, we should compute the zeros $x_k$ of $u$ in the
interval $]\delta ,d-\delta [$ and define the frequencies:
\begin{equation}\label{eq:freqid}
\nu_k~=~\begin{cases} c\beta\sqrt{(x_{k+1})^{-2}
+(d-x_{k+1})^{-2}} ~~~& \forall x\in ]x_k, x_{k+1}[~~~x_{k+1}\leq d/2 \\  \\
c\beta\sqrt{(x_{k})^{-2}
+(d-x_{k})^{-2}} ~~~& \forall x\in ]x_k, x_{k+1}[~~~x_{k}\geq d/2
\end{cases}
\end{equation}
which are an average of the frequencies induced by each plate
separately. Again, we put a limit to $\nu_k$ according to the
inequality in (\ref{eq:limite}). In practice, we will not consider
the indices such that $\nu_k\geq\nu_{\rm max}$, in the global
computation of the energy.
\par\smallskip

\begin{center}
\begin{figure}[!h]
\centerline{\includegraphics[width=13.cm,height=4.cm]{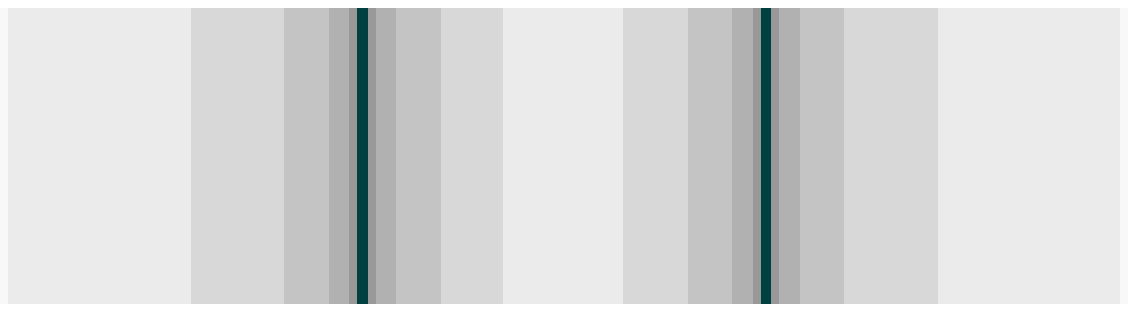}}
\centerline{\includegraphics[width=13.cm,height=4.cm]{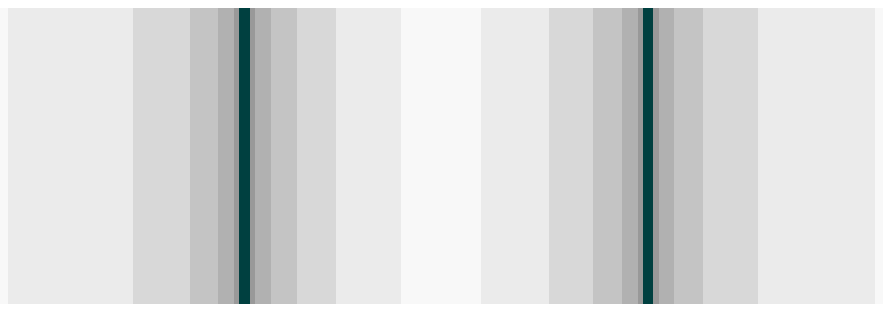}}
\begin{caption}{\small Stratification of electromagnetic layers between and
outside two conductive surfaces at a given distance $d$ (the
vertical dimension is not in scale). The picture on top is related
to a value of $\lambda$ which is smaller than that corresponding
to the picture on bottom. A fractal pattern is obtained by
superimposing many pictures as the above ones, by varying
$\lambda$ in the range of the eigenvalues in (\ref{eq:eige}).}
\end{caption}
\end{figure}
\end{center}
\vspace{-.4cm}

We expect a distribution of the bands as in figure 5, where one
can clearly notice a reduced energy in the central part. Now, we
would like to quantify the missing contribution at the interior.
Near the plates, independently of the side, the width of the bands
is practically the same, i.e., the solutions of (\ref{eq:lapotem})
and (\ref{eq:lapotem2}) behave in the same way. Therefore, we are
interested to know the displacement of the zeros of $u$ in
(\ref{eq:lapotem2}) in the central part of the interval $]0,d[$.
\par\smallskip

For $x\in [\delta ,d-\delta ]$, due to the fact that we now have
two boundaries, we need to make additional assumptions. Actually,
we would like to impose for instance $u(\delta )=u(d-\delta )=0$
(grounded plates), but these conditions produce in general only
the trivial solution $u\equiv 0$. However, for certain values of
$d$, the differential operator in (\ref{eq:lapotem2}) (whose sign
is not definite) is allowed to admit eigenfunctions corresponding
to the zero eigenvalue. Therefore, one can check that there is a
sequence of values $d$ such that problem (\ref{eq:lapotem2}) has
nontrivial solutions even when $u (\delta )=u(d-\delta )=0$. We
claim that, in correspondence to these values, the set of layers
is well-defined and somehow in equilibrium. For other choices of
$d$, some layers (in particular the central ones) still do not
possess the right energy to form a complete photon. In these
unclear situations we are unable to predict, with strict accuracy,
the size of the
various bands. Nevertheless,  we can argue as follows. Suppose
that for a given $d$, we can find a nontrivial eigenfunction $u$
and compute its zeros. Then, as $d$ gets larger we obtain an
intermediate situation where the global energy is going to be
greater than the previous one. By increasing $d$ again, further
electromagnetic energy is captured between the plates, until we
reach another stable configuration where $u$ turns out to be the
successive eigenfunction of (\ref{eq:lapotem2}). The new $u$
displays more oscillations, and, for this reason, realizes a
greater quantity of layers.
\par\smallskip

In figure 6, we see the plot of an  eigenfunction $u$, obtained
for a suitable value of $d$. In this case, $u$ is an odd function
vanishing for $x=d/2$. Therefore, together with the conditions $u
(\delta )=u(d-\delta )=0$, we also get $u^\prime (\delta
)=u^\prime(d-\delta )$. In truth, $u$ is determined up to a
multiplicative constant, but, of course, this has no effect on the
zeros. As we said before, the number of oscillations of $u$ in
$]\delta , d-\delta [$ depends on the magnitude of $d$. Therefore,
by slowly increasing the size of $d$, one can observe the
transition from a state with $n$ nodes to that with $n+2$ nodes,
corresponding to the creation of a new central layer, through the
absorbtion of a lower-frequency photon. A few steps of this
evolution can be viewed in figure 7. The situation is very similar
when we impose $u (\delta )=u(d-\delta )$ (plates at the same
potential) together with  $u^\prime (\delta )=u^\prime(d-\delta
)=0$ (uncharged plates). In this case the function $u$ is even.
\par\smallskip

The detection of a quantized system of distances, realizing sets
of completed shells (each shell carries the minimum allowed energy
and there is no energy in excess to form a further shell), may
explain the various states of excitation of an atom. As customary,
the transition from one state to another is justified by the
release or the absorbtion of photons. If this hypothesis is
correct, one should be able to explain the origins of the
quantized atomic structure. Numerical 3-D experiments are
currently under way in order to understand if this guess is
realistic.
\par\smallskip

\begin{center}
\begin{figure}[!h]
\centerline{\includegraphics[width=6.8cm,height=6.8cm]{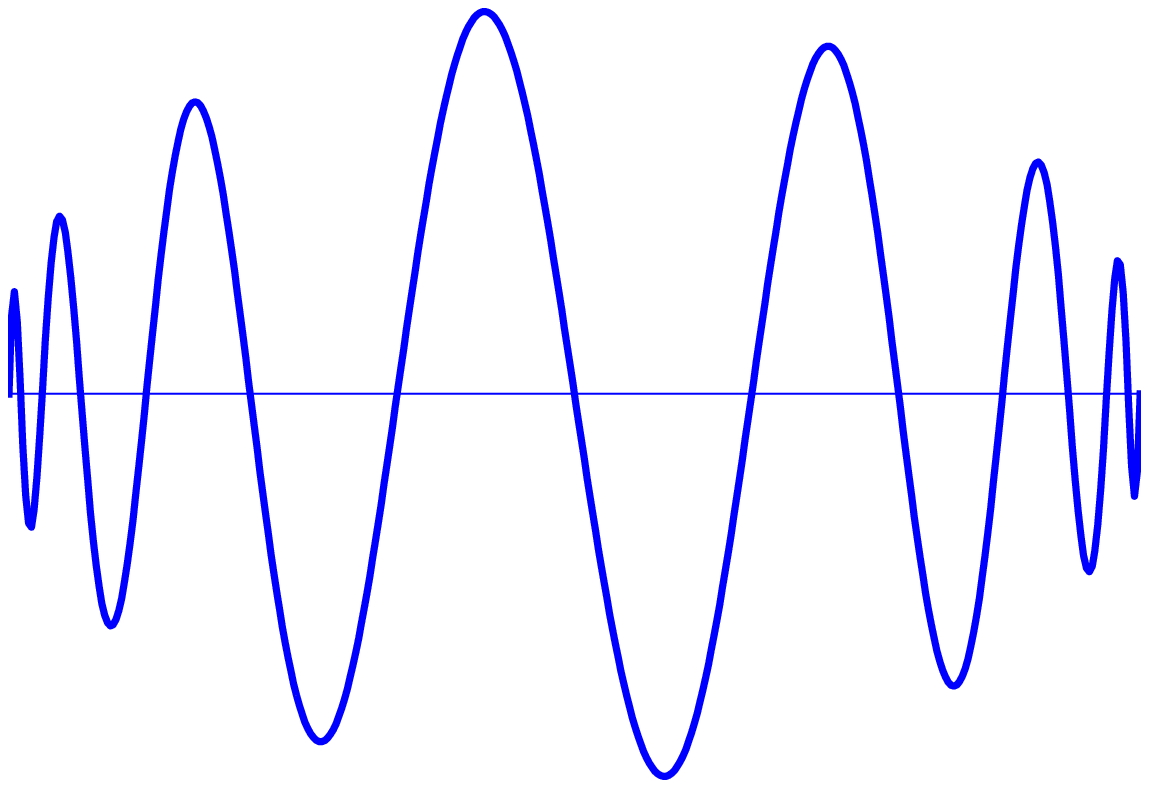}
\hspace{-.3cm}\includegraphics[width=6.8cm,height=6.8cm]{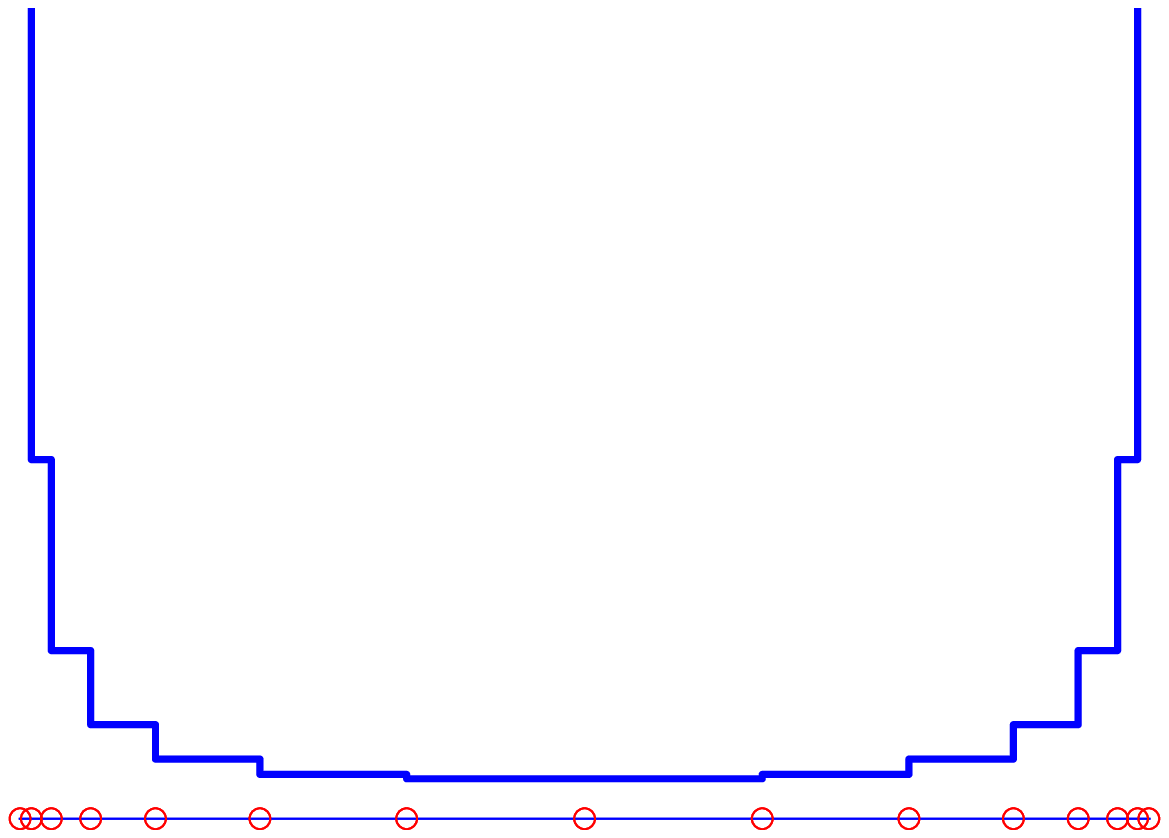}}
\begin{caption}{\small Qualitative plot of the function $u$, solution
of (\ref{eq:lapotem2}), with $u (\delta )=u(d-\delta)=0$,
and the corresponding energy built on its zeros (represented by the dots).}
\end{caption}
\end{figure}
\end{center}

\begin{center}
\begin{figure}[!h]
\centerline{\includegraphics[width=8.cm,height=6.cm]{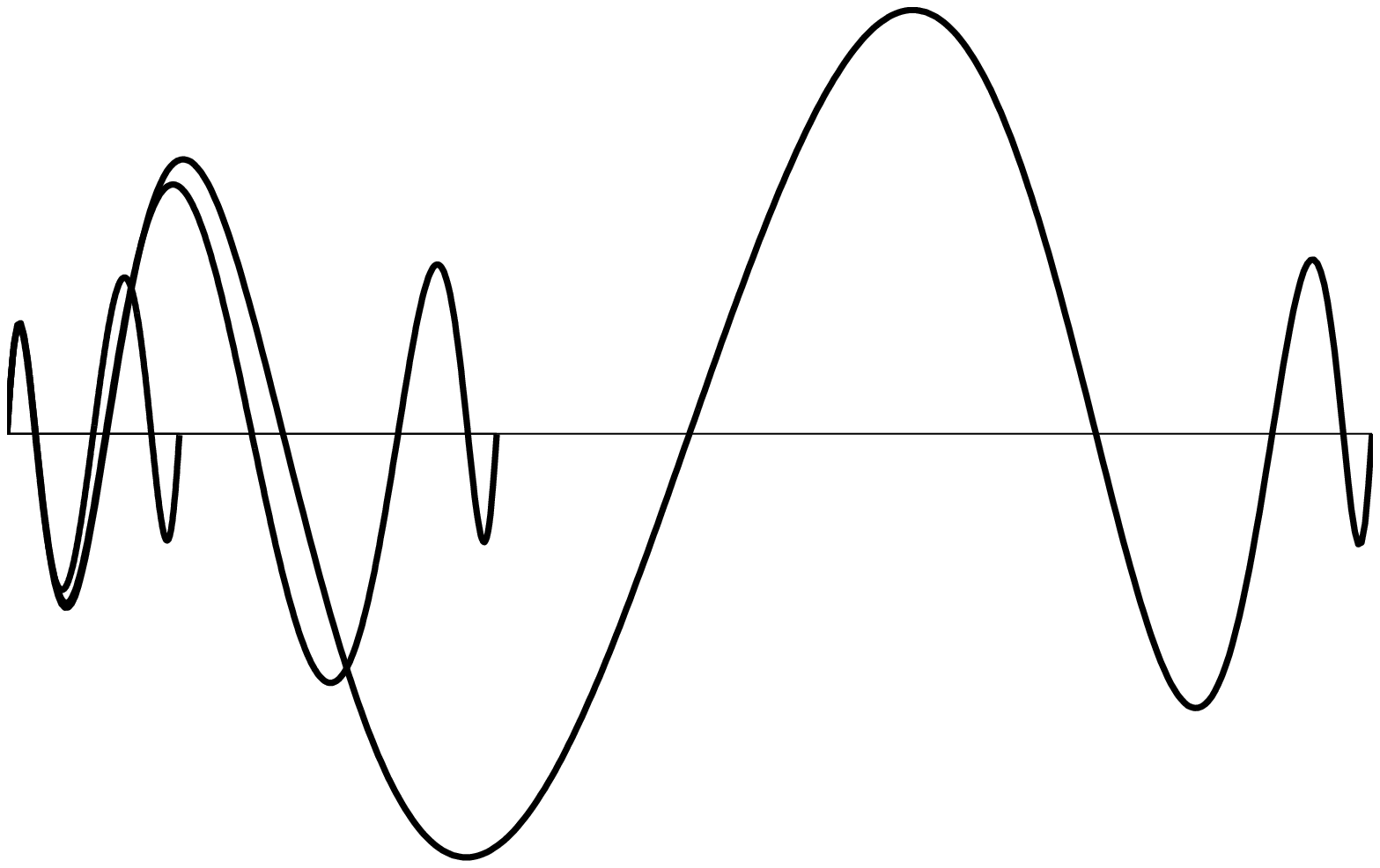}}\vspace{.5cm}
\centerline{\includegraphics[width=8.cm,height=6.cm]{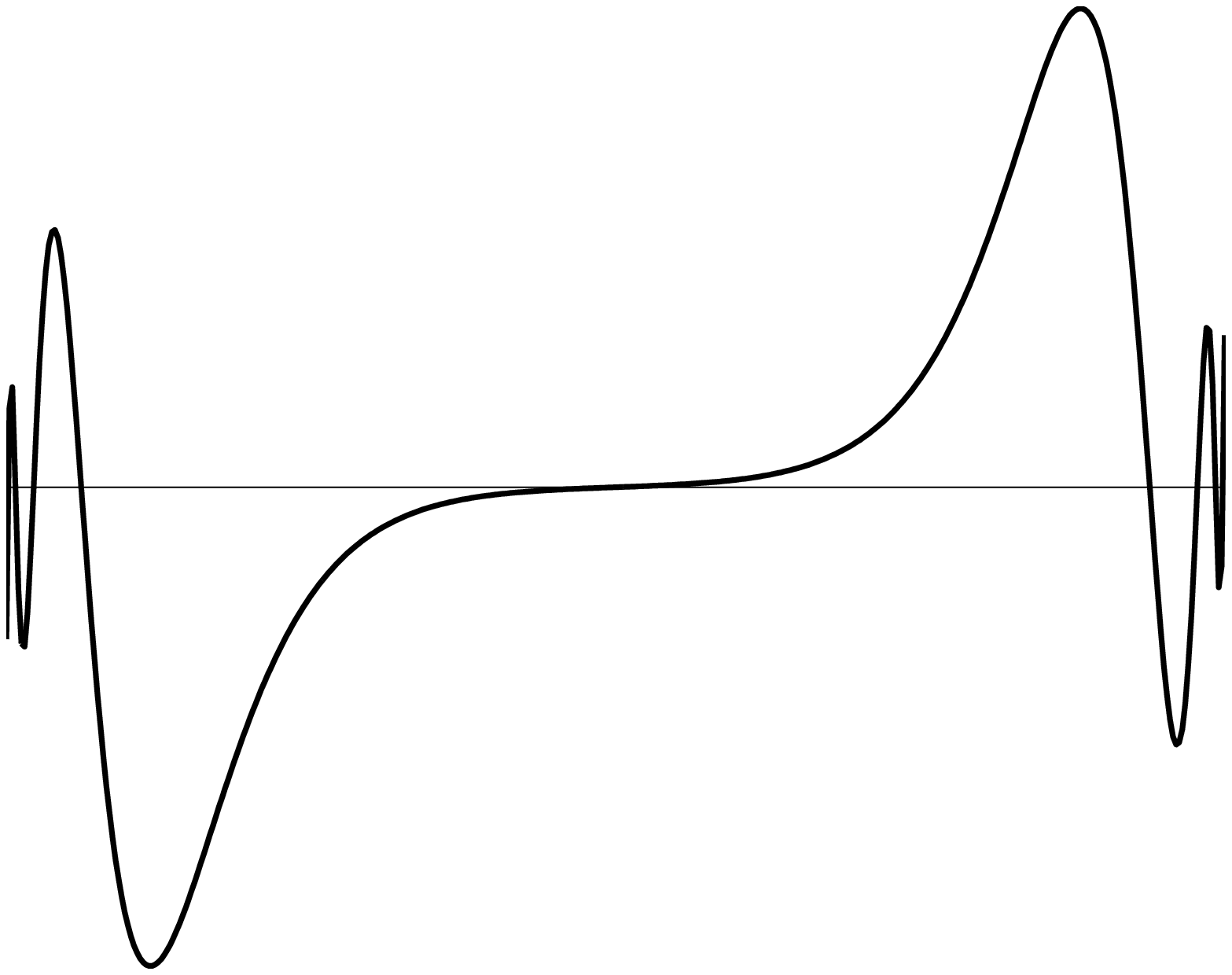}}
\begin{caption}{\small Solutions of (\ref{eq:lapotem2}) for different values
of the parameter $d$. As $d$ grows, more oscillations appear, until, depending
on the ratio $\beta /\lambda$, one reaches a complete separation.
In the picture on top the various $u$ are actually eigenfunctions with zero boundary
conditions. This is not true for the picture on bottom (not in
the same scale-length as the above ones), because the plates are
too far apart and the energy in the middle is too low
to allow the formation of an entire photon.}
\end{caption}
\end{figure}
\end{center}
\vspace{-.5cm}


We wish to show that the energy of the band  in the middle  (the
one centered at $x=d/2$) is proportional to $d^{-2}$. We can
attempt a first approximate computation by taking $\lambda =0$ in
(\ref{eq:lapotem2}) (note, for example, that $\lambda$ is
negligible when the plates are very close to each other).
Afterwards, following the trick used in section 4, we evaluate a
primitive of $f$:

$$
F~=~\beta\left[\sqrt{2}~{\rm arcsinh}\hspace{-.1cm}\left(\frac{2x}{d}-1\right)+
{\rm arctanh}\hspace{-.1cm}\left(\frac{d-x}{\sqrt{d^2-2xd+2x^2}}\right)\right.-~~~~~~~~~~
$$
\begin{equation}\label{eq:primi2}
~~~~~~~~~~~~~~~~~~~~\left.{\rm arctanh}\hspace{-.1cm}\left(\frac{x}{\sqrt{d^2-2xd+2x^2}}
\right)\right]
\end{equation}
and, successively, we compute the the values $x_k$ such that
$F=k\pi$, where $k$ is an integer.


\par\smallskip
For $k=0$ one has $x_k=d/2$ and $F(x_k)=0$. Hence, if we
want to know what happens at the center of the interval $]0,d[$,
we need to look at the distance $\vert x_1 -x_{-1}\vert$.
Near $x=d/2$, in the first approximation, one can write:
$$F(x)\approx F(\textstyle{\frac{d}{2}})+F^\prime(\textstyle{\frac{d}{2}})
(x-\textstyle{\frac{d}{2}})=f(\textstyle{\frac{d}{2}})(x-\textstyle{\frac{d}{2}})
=\beta\sqrt{2}~(\textstyle{\frac{2}{d}}x-1)
$$
Therefore, we get $~F(x)=\pi$ when $~x\approx\frac{1}{2}d(1+\pi
/\beta\sqrt{2})$, which means that the quantity $\vert x_1
-x_{-1}\vert$ is proportional to $d$.
\par\smallskip

Based on (\ref{eq:freqid}), the frequency in $d/2$ is approximately
$\nu_{\rm central}=2c\beta /d$, so that, using (\ref{eq:eneri}), as far as the
the central band is concerned, the energy per unit of volume is:
\begin{equation}\label{eq:enerix2}
{\cal E}_{\rm central} ~=~\frac{h\nu_{\rm central}}{2A~\vert x_1-x_{-1}\vert
}~\approx ~\frac{K}{d^2}
\end{equation}
where $K$ is a constant.
\par\smallskip

Outside the plates, there are certainly more bands.  Reasonably,
the exceeding ones carry more than the energy ${\cal E}_{\rm
central}$. In addition, considering (\ref{eq:enerix}), let us
suppose that we can find $k_0$ such that ${\cal E}_{k_0}={\cal
E}_{\rm central}$. Then, due to the exponential growth of the
nodes $x_k$, the sum $\sum_{k=k_0}^{+\infty} {\cal E}_k$ is
proportional to ${\cal E}_{\rm central}$ (use the relation:
$\sum_{m=m_0}^{+\infty} \sigma^m =\sigma^{m_0}/(1-\sigma )$, valid
for $\vert\sigma\vert <1$). Therefore, due to (\ref{eq:enerix2}),
the missing energy behaves as $d^{-2}$.
\par\smallskip

By differentiating the energy difference with respect to $d$, one
gets that the force is proportional to $-1/d^3$. This result does
not agree with formula (\ref{eq:casa}). Nevertheless, we do not
have to forget that the estimates obtained above, should be
referred to a single specific choice of $\lambda$. Indeed, we have
a multiplicity of layer structures, depending on the different
values in (\ref{eq:eige}). We recall that, by increasing
$\lambda$, we produce a compression of the band range around the
plates (see figure 5), that have a global extension within the
interval  $[\delta ,\beta /\lambda ]$.
\par\smallskip

The whole phenomenon is explained as follows.  When the plates are
far apart (at distance greater than $\beta \sqrt{A}$), they are
practically independent and no forces are exerted. As the distance
is reduced, some central bands are eliminated by photon emission
and an attractive force proportional to $d^{-3}$ starts acting on
the plates. The effect is due to the activation of the lowest
transversal mode ($n=m=1$ in (\ref{eq:eige})). The other higher
modes determine the presence of bands that are still not involved.
As we further reduce the distance $d$, together with the
elimination of some layers corresponding to the smallest
$\lambda$, other layers (those for $n,m\leq 2$) are successively
discarded, increasing the strength of the force, becoming twice
the previous one. Of course, the process is not continuous, but
follows a quantized path. The emission of photons is subjected to
complicated rules (we remind that there is a fractal structure
underneath), so it might be not easily detected. Moreover, for
simplicity, we threw away the time variable, but we do not have to
forget that, in reality, the dynamics is extremely complex.
Asymptotically, when $d$ gets really small, the number of systems
of layers involved is proportional to $1/d$. In fact, all the
eigenvalues corresponding to $\lambda <\beta /d~$ ($n=m\approx
d^{-1}$) are associated with bands that are involved in the
elimination process. Since the band systems are all similar (one
can be mapped into another via a linear transformation), the total
force is proportional to $d^{-3}$ multiplied by $d^{-1}$, finally
yielding (\ref{eq:casa}). When $d$ is close to $\delta$ the plates
are practically touching.
\par\smallskip

Our analysis reveals a non uniform spectrum of frequencies.
Weighted spectra, mainly chosen according to statistics arguments
and bringing to finite energy sums, may be introduced. A few
papers dealing with this filtering procedure are for instance:
\cite{ford}, \cite{ford2}, \cite{ellingsen}. The emission of
photons, when the configuration of the plates is nonstationary, is
a known phenomenon (see for instance \cite{moore} and
\cite{dodonov}), often mentioned as dynamical Casimir effect. The
present paper may help to clarify (also providing a mathematical
model) the reasons for this amazing photon creation from vacuum.
In \cite{bak}, interesting connections linking the ``critical
states'' in natural systems and the production of the so called
{\sl pink noise}, are made. We suspect that there is an analogy
between the results in \cite{bak} and the states of equilibrium of
the plates, corresponding to the eigenfunctions of problem
(\ref{eq:lapotem2}), where $\lambda$ is an eigenvalue related to
the transversal modes. The relative movement of the plates passes
through the emission of a complicated sequence of photons
exhibiting a well-determined spectrum of frequencies, that could
be put in relation with some specific type of noise. However, the
analysis of this aspect is out of the scopes of the present paper.
\par\smallskip

At the moment, little can be said concerning the proportionality
constant in (\ref{eq:casa}). The Planck constant and the speed of
light are certainly involved, since they are used to set the
energy levels at the ground-state. We do not advance any
hypothesis concerning the values of $\beta$ and $\delta$. The
first one acts on the rate of growth of the layers, that follows a
geometrical rule. The second one has effect on the global amount
of energy. Some information could be recovered from the vast
literature on power-law scaling for metals (see for example
\cite{ma}).
\par\smallskip

This explanation of the Casimir effect may sound quite involved
(on the other hand, the usual explanations looks rather
simplistic), but agrees with an electrodynamic interpretation of
the universe that could be spent to clarify many other facts,
or, perhaps, to extract energy from the vacuum (\cite{cole}, \cite{bearden}).

\par\medskip
\setcounter{equation}{0}
\section{Other configurations}

We can now charge the two plates with opposite sign. Everybody
would tell us that a stronger attraction is felt. But, how this
phenomenon really happens? How can we justify the action at a
distance of the Coulomb law? Our explanation is that the same
identical layers, developing in the uncharged case, are also
present in this case, transferring the information from one plate
to the other. As argued in \cite{funarol}, the peculiar
geometrical structure of the layers is decided by the
time-dependent part of the electromagnetic fields (always
existing, even at zero temperature), while the stationary part is
just added to it. The stationary part emphasizes the negative
pressure, resulting in a more pronounced suction effect.
\par\smallskip




In the vicinity of the plates the steady electric field is almost
constant, and tends to zero far from them. As done for the
treatment of the uncharged plates, by measuring the internal and
the external energy densities $~\frac{1}{2}\epsilon_0\vert {\bf
E}\vert^2$, we come out with a resulting force pushing from
outside, which is responsible for the attraction. Such a
computation is trivial. However, here we are proposing an alternative
viewpoint by arguing that the evaluation of the energy is not the
integral of a continuous functions, but the sum of the partial
contributions of the various layers. The terms of the sum combine
the stationary field, with the time-dependent fields of the layers
studied in the previous sections. If we are not at atomic scale,
the second field is negligible with respect to the first one, so
that the continuous stationary component dominates, masking the
quantum effects.
\par\smallskip


Another issue we would like to discuss is the possibility of
considering different geometrical assets of the plates. A typical
example is the one of two spherical uncharged conductive plates at
short distance $d$. The results about this case are controversial.
From one side,  with very technical arguments, in \cite{boyer} it
is shown that, surprisingly, the plates should repel, i.e., the
inside zero-point energy is greater that that outside. It seems
however that there are no documented laboratory tests in favor of
this hypothesis, mainly due to the difficulty of setting up the
experiments. A general overview, together with a new approach to
the theoretical treatment of hemispheric surfaces, are provided in
\cite{cho} and \cite{ozcan}. In recent experiments (see \cite{munday},
\cite{lisanti}), within the context of suitable geometrical environments at
nanoscale level, the detection of repelling Casimir forces has
been observed. Other extensions and a good list of
references are provided for instance in \cite{farina}.

\par\smallskip

From our viewpoint the situation is delicate. We recall that,
according to our theory, the electromagnetic energy, pervading the
universe from the very beginning, is organized, by the rapidly
spinning particles inside a molecular texture, to follow
well-determined patterns. This imprinting specifically depends on
the type of material under study and its geometrical properties.
As far as we are concerned, at the nanoscale, all the forces
(Coulomb, van der Waals, Casimir, and even gravitational) are
manifestation of the same modelling equations, combining in the
appropriate manner (with a dosage depending on the context)
electromagnetic phenomena with the space-time
geometry, via Einstein's equation.
\par\smallskip

The case of the plane parallel metallic plates seems to be
reasonably covered by the analysis developed here (provided we
neglect what happens at their boundaries), because it displays
strong symmetry properties. Nevertheless, we had to choose a
minimum width $\delta$ to give a bound to the infinitesimal
fractal structure encountered when approaching matter (otherwise
atoms would be indefinitely small). But, what happens when a plate
is bent? On one side the molecular structure is stretched, on
the other is compressed. In addition, this process may alter
temperature. This means that, unless we have a very good knowledge
of the behavior of matter at extremely short distances, it is hard
to predict the shape of the electromagnetic layers circulating
around. We also know that the innermost layers are those
displaying the highest energy, influencing in this way the
magnitude and the frequency of the successive layers. In the end, with
some appropriate assumptions on the properties of matter, an
analysis of the Casimir effect for spherical geometries (for
instance) could be carried out with the help of equations
(\ref{eq:lapote}) and (\ref{eq:laposep}). However, for the moment,
this is an issue we would prefer to avoid.

\par


\medskip

\begin{thebibliography}{99}

\bibitem{bak} Bak P., Tang C.,  Wiesenfeld K. (1988), Self-organized criticality,
Phys. Rev. A, {\bf 38}, pp. 364-374.

\bibitem{barabasi} Barab\'asi A. -L., Stanley H. E. (1995), {\sl Fractal Concepts in
Surface Growth}, Cambridge Univ. Press.

\bibitem{bearden} Bearden T., Bedini J. (2006), {\sl Free Energy Generation, Circuits
\& Schematics}, Cheniere Press.


\bibitem{bradley} Bradley J. -C., Chen H. -M., Crawford J., Eckert J., Ernazarova K.,
Kurzeja T., Lin M., McGee M., Nadler W., Stephens S. G. (1997), Creating electrical contacts
between metal particles using directed elecrochemical growth, Nature, {\bf 389},
pp. 268-271.

\bibitem{boyer} Boyer T. H. (1968), Quantum electromagnetic zero-point energy of a
conducting spherical shell and the Casimir model for  a charged particle, Phys. Rev.,
{\bf 174}, n. 5, pp. 1764-1776.

\bibitem{casimir} Casimir H. B. G. (1948), On the attraction between two perfectly
conducting plates, Proc. Kon. Nederland. Akad. Wetensch., {\bf B51}, p. 793.

\bibitem{casimir2} Casimir H. B. G., Polder D. (1948), The influence of retardation on the
London-van der Waals forces, Phys. Rev., {\bf 73}, pp. 360-372.

\bibitem{chinosi} Chinosi C., Della Croce L., Funaro D., Rotating electromagnetic waves
in toroid-shaped  regions - part I, submitted.

\bibitem{cho} Cho S. N. (2008), Is repulsive Casimir force physical?, arXiv:quant-ph/0408184v1

\bibitem{cole} Cole D. C., Puthoff H. E. (1993), Extracting energy and heat from the vacuum,
Phys. Rev. E, {\bf 48}, n. 2, pp. 1562-1565.

\bibitem{dodd} Dodd J. N. (1991), {\sl Atoms and Light: Interactions}, Springer.

\bibitem{dodonov} Dodonov V. V., Dodonov A. V. (2005), Quantum harmonic oscillator and
nonstationary Casimir effect, J. of Russian Laser Research, {\bf 26}, n. 6, pp. 445-483.

\bibitem{ellingsen} Ellingsen S. A. (2008), Frequency spectrum of the Casimir force:
interpretation and a paradox, EPL, {\bf 82}, 53001.

\bibitem{farina} Farina C. (2006), The Casimir effect: some aspects, Brazilian J. of Physics,
{\bf 36}, n. 4A, pp. 1137-1149.

\bibitem{ford} Ford L. H. (1988), Spectrum of the Casimir effect, Phys. Rev. D, {\bf 38},
n. 2, pp. 528-532.

\bibitem{ford2} Ford L. H. (2007), Frequency spectra and probability distributions for
quantum fluctuations, Int. J. Theor. Phys., {\bf 46}, pp. 2218-2226.

\bibitem{funarol} Funaro D. (2008), {\sl Electromagnetism and the Structure of Matter},
World Scientific, Singapore.

\bibitem{funarow} Funaro D. (2008), http://cdm.unimo.it/home/matematica/funaro.daniele\-
/phys.htm

\bibitem{ibison} Ibison M., Haisch B. (1996), Quantum and classical statistics of the
electromagnetic zero-point field, Phys. Rev. A, {\bf 54}, n. 4. pp. 2737-2744.

\bibitem{lisanti} Lisanti M., Iannuzzi D., Capasso F. (2005), Observation of the
skin-depth effect on the Casimir force between metallic surfaces, PNAS, {\bf 102},
n. 34, pp. 11989-11992.

\bibitem{loudon} Loudon R. (2000), {\sl The Quantum Theory of Light}, Oxford Univ. Press.

\bibitem{ma} Ma D., Stoica A. D., Wang X. -L. (2009), Power-law scaling
and fractal nature of medium-range order in metallic glasses,
Nature Materials, {\bf 8}, pp. 30-34.

\bibitem{mandelbrot} Mandelbrot (1982), {\sl The fractal Geometry of Nature}, Freeman,
San Francisco.

\bibitem{marani} Marani M., Banavar J. R., Caldarelli G., Maritan A.,
Rinaldo A. (1998), Stationary self-organized fractal structures in an open,
dissipative electrical system, J. Phys. A: Math. Gen., {\bf 31}, L337.

\bibitem{milonni} Milonni P. W. (1993), {\sl The Quantum Vacuum: an Introduction to
Quantum Electrodynamics}, Academic Press.

\bibitem{moore} Moore G. T. (1970), Quantum theory of the rlectromagnetic field in
a variable-length one-dimensional cavity, J. Math. Phys., {\bf 11}, n. 9, p. 2679.

\bibitem{munday} Munday J. N., Capasso F. (2007), Precision measurements of the
Casimir-Lifshitz force in a fluid, Phys. Rev. A, {\bf 75}, 060102(R).

\bibitem{munday2} Munday J. N., Capasso F., Parsegian V. A. (2008), Measured long-range
repulsive Casimir-Lifshitz forces, Nature, {\bf 457}, pp. 170-173.

\bibitem{ozcan} \"Ozcan M. (2005), Casimir energy between two concentric half
spheres, Phys. Letters A, {\bf 344}, pp. 307-315.

\bibitem{peitgen} Peitgen H. -O., Hartmut J., Dietmar S. (1992),
{\sl Chaos and Fractals, New Fontiers of Science}, Springer, New
York.

\end{thebibliography}
\end{document}